\documentclass[a4 paper, 11pt,twoside]{article}
\usepackage{comment}

\usepackage{amsmath,amssymb}
\usepackage{lipsum}
\usepackage{hyperref}
\usepackage{physics}
\usepackage{comment}
\usepackage{feynmf}
\usepackage{graphicx}
\usepackage{mdframed}
\usepackage{endfloat}
\usepackage{fullpage}                           % Makes the page margins smaller to a predefined one.
\usepackage[lmargin=0.6in,rmargin=0.6in,tmargin=0.6in,headsep=.2in]{geometry}
\usepackage{graphicx}
\usepackage{forloop}
\usepackage{etoolbox}
\usepackage{calc}
\usepackage{wrapfig,lipsum,booktabs} % To produce the empty space on the left
\usepackage{xtab} % To produce the empty space on the left
\usepackage[dvipsnames]{xcolor}
\usepackage[normalem]{ulem}
\usepackage{sectsty}% http://ctan.org/pkg/sectsty
\allsectionsfont{\color{Brown}}
\makeatletter
\def\@seccntformat#1{\@ifundefined{#1@cntformat}%
{\csname the#1\endcsname\;}%  default
{\csname #1@cntformat\endcsname}% individual control
}
\def\section@cntformat{\thesection.\;} % Dot after the section number
\def\subsection@cntformat{\thesubsection.\;} % Dot after the subsection number
\makeatother
\usepackage{hyperref}
\hypersetup{
    colorlinks=true,
    urlcolor=blue,
    linkcolor=black,
    citecolor = black}

\usepackage{fancyhdr}
\begin{document}
%\begin{CJK*}{}{}
%\affiliation{fiziSi/Bijm LLC \\  2245 Hillsbury Rd. \\Westlake Village, CA 91361 \\ bijan.berenji@fizisim.com }
%\affiliation{SLAC National Accelerator Laboratory \\ 2575 Sand Hill Rd., Menlo Park CA 94025 \\ bijanb@slac.stanford.edu}
%\maketitle
	{\small
%  ---------------   Normal Headers  ------------------
\fancyhf{} % clear all fields
\fancyhead[LE,RO]{\textsf{Bijan Berenji}}
%\fancyhead[C]{Paper Title: Running Title}
%\fancyfoot[LO,RE]{\includegraphics[scale=0.12]{index1.png}}
\fancyfoot[LE,RO]{\thepage}
\renewcommand{\headrulewidth}{1pt} % to remove line on header
\renewcommand{\footrulewidth}{1pt} % to remove line on footer
% ------------------ Header for first page -----------------
\fancypagestyle{first}{%
  \fancyhf{}% clear all header and footer fields
  \fancyfoot[L]{{\footnotesize \textsf{DOI: \href{10.4236/***.2024.*****}{\color{blue}\uline{10.4236/***.2024.*****}} $\;$**** **, 2024}}}%
  \fancyhead[R]{{\bf\small \textsf{Journal of ****, 2024, *, *-*}}\\
\href{http://www.scirp.org/journal/***}{\color{blue}\uline{\textsf{http://www.scirp.org/journal/***}}}\\
\textsf{ISSN Online:****-****}\\
\textsf{ISSN Print:****-****}}%
  \renewcommand{\headrulewidth}{0pt}% to remove line on header
  \renewcommand{\footrulewidth}{0pt}% to remove line on footer
}
%\begin{document}
%\thispagestyle{first}
%\vspace*{3cm}
%%%%%%%%%  TITLE %%%%%%%%%%%%%%%%%
{\noindent\huge\bf Axion Gamma-Ray Signatures from Quark Matter in Neutron Stars and Gravitational Wave Comparisons}
\hspace{5.0cm}\vspace{5.0cm}
%%%%%%%%%%%%%%%%  Author Data %%%%%%%%%%%%%%%%%%%
\vspace{5.0cm} 
\
\\
{\bf\large Bijan Berenji}\\
{\bf \large SLAC National Accelerator Laboratory, Menlo Park, CA 94025 USA}\\
{\bf\large Email: bijanb@slac.stanford.edu}\\
%%%%%%%%%%%   The Information Bar on the Left %%%%%%%%%%%
\begin{wraptable}{l}{5.1cm}
{\footnotesize
\begin{xtabular*}{0.3\textwidth}{p{5cm}}
\url{http://dx.doi.org/10.4236/***.2024.*****}\\
{\bf Received: **** **, ***}\\
{\bf Accepted: **** **, ***}\\
{\bf Published: **** **, ***}\\
Copyright \copyright$\;$2024 by author(s) and Scientific Research Publishing Inc.\\
This work is licensed under the Creative Commons Attribution International License (CC BY 4.0).\\
\url{http://creativecommons.org/licenses/by/4.0/}\\
%\includegraphics[width=2.5cm,height=0.72cm]{ccby40.png}$\;$\includegraphics[width=2.5cm,height=0.75cm]{openaccess.jpg}\\
%%%%%%%%%%%%%%%%%%% IMPORTANT FOR AUTHORS %%%%%%%%
% Copy and paste the following line depending on the number of full pages in your document (i.e. before using the template)
%\color{white}\lipsum[1-60]}% If the number of pages is less than 15
%{\color{white}\lipsum[1-60]}% uncomment if the number of pages is more than 15 and less than 30
%{\color{white}\lipsum[1-60]}% uncomment if the number of pages is more than 30 and smallless than 45
%%%%%%%%%%%%%%%%%%%%%%%%%%%%%%%%%%%%%%%%%%%%%%%%%%%%%
\end{xtabular*}
}
\end{wraptable}

\pagebreak
{\color{Brown}\rule{0.7\textwidth}{2pt}\\}
{\color{Brown}\bf\large Abstract}\\\color{Black}
%{\color{Brown}\rule{0.7\textwidth}{2pt}}\\[0.2cm]
%{\color{Brown}\bf\large Abstract}\\
We present a theoretical model for detecting axions from neutron stars in a QCD phase of quark matter.  The axions would be produced from a quark-antiquark pair $u\bar{u}$ or $d\bar{d}$, in loop(s) involving gluons. The chiral anomaly of QCD and the spontaneously broken symmetry are invoked to explain the non-conservation of the axion current.  From the coupling form factors, the axion emissivities $\epsilon_a$ can be derived, from which fluxes can be determined.   We predict a photon flux, which may be detectable by Fermi LAT, and limits on the QCD mass $m_a$.  In this model, axions decay to gamma rays in a 2-photon vertex. We may determine the expected fluxes from the theoretical emissivity. The sensitivity curve from the Fermi Large Area Telescope (Fermi LAT)  would allow axion mass constraints for neutron stars as low as $m_a \le 10^{-14}$ eV 95$\% C.L.$.  Axions could thus be detectable in gamma rays for neutron stars as distant as 100 kpc.  A signal from LIGO GWS 170817 could be placed from the NS-NS merger, which gives an upper limit of $m_a \le 10^{-10}$ eV.
	\vspace{0.5cm}\\
{\color{Brown}\bf\large Keywords}\\
astrophysics; phenomenology; QCD axion; neutron stars; nuclear theory; gamma rays; gravitational waves; Fermi-LAT
\vspace{0cm}\\
{\color{Brown}\rule{0.7\textwidth}{2pt}}

%\end{frontmatter}
%\maketitle

\newcommand{\ra}{q^2}
%\newcommand{\q^4}{q^4}
%\begin{document}
\section{Introduction} 

The QCD axion has been investigated from its production in nucleon-nucleon bremsstrahlung in supernovae or neutron stars.  Many models of neutron stars include QCD phases~\cite{Ruster2005neutrino,Hassaneen,Hujeirat}, with quark matter in addition to hadronic phases.   The assumed temperature may be in the range 10 MeV - 100 MeV, which may be possible under certain conditions.  For example, In a binary neutron star merger, the conditions exist for a QCD phase transition, during which the temperature of the neutron star core would rise to between 50 to 90 MeV~\cite{chenNSmerger}.  In this situation, axion production could start at a high flux, generating gamma-rays between 30 MeV to $\simeq 400 MeV$. In our recent work Refs.~\cite{berenjiAxions,extAxion} we have described the model for the point-source and extended-source emission of axions and their corresponding gamma-ray signals from neutron stars.  It may be possible to detect axions or to set more stringent limits on the axion mass immediately following a gravitational wave signal detected by LIGO/Virgo.  The continuous gravitational wave signature of a neutron star merger has been discussed in Ref~\cite{continuousGW}.  It has also been demonstrated that emission of Kaluza Klein gravitons could occur in a supernova, which would produce a gamma-ray signature~\cite{hanhart1987}.  The conditions, i.e., the temperature,  would be very similar to that of a neutron star in a QCD phase transition.    Previous limits on Large Extra Dimensions with Kaluza-Klein gravitons have been placed using a model for neutron stars with Fermi-LAT observations~\cite{LEDFermiJCAP}.  The gamma-ray signatures in both cases could be detected by Fermi LAT.  

The axion is a well-motivated particle of theoretical physics.  This light pseudoscalar boson arises as the pseudo Nambu-Goldstone boson of the spontaneously broken $U(1)$ Peccei--Quinn symmetry of quantum chromodynamics (QCD), which explains the absence of the neutron electric dipole moment~\cite{axionBook,chengLi}, and thereby solves the strong $CP$ problem of particle physics~\cite{PQ,wein,wilczek}.   In addition, it is a possible candidate for cold dark matter~\cite{preskill1983cosmology}.  Astrophysical searches for axions generally involve constraints from cosmology or stellar evolution~\cite{raffelt,raffeltStellar}.  Many astrophysical studies placing limits on the axion mass have also considered axion production via photon-to-axion conversion from astrophysical and cosmological sources such as type Ia supernovae and extra-galactic background light~\cite{horns2012hardening,sanchez2009hints,brockway1996sn,csaki2002dimming}.   Searches for axions from neutron stars, have generally considered the axion-nucleon coupling, but we may now consider the axion-quark-antiquark coupling. 	Axions are deeply related to QCD, which refers to the dynamics of the strong force.   Asymptotic freedom refers to the observation that the force between particles decreases asymptotically as the energy scale increases and the corresponding length scale decreases.  Color confinement refers to the interaction force between two color charges remains constant as they are separated.   Another notable feature of QCD is renormalization, the strength of the interaction running with the energy scale. 

	%We plan to set bounds on the axion mass $m_a$ by considering radiative decays of axions produced by nucleon-nucleon bremsstrahlung in neutron stars.   

Extended gamma-ray sources have been extensively studied with the Fermi LAT~\cite{lande2012search}, including pulsar wind nebulae and supernova remnants.    In addition, dark matter in galaxies may be modeled as extended sources of gamma rays~\cite{AbazajianDM}.  We may note that spatially-extended emission from axions may occur in the vicinity of supernova remnants.  Decays that occur at a distance from supernova remnants have been considered in Ref.~\cite{giannotti}.  Here, we consider variation on the point-source neutron star model considered previously, and consider extended emission due to axions decaying at a certain distance away from the source.  It has been shown in Ref.~\cite{BerenjiAAS2017,BerenjiConstraints1} that axion decay may render the neutron star as an extended source of $\simeq 1^\circ$ radius in the sky for a source of $d\lesssim 300$ pc.  

    Several gravitational wave events have been detected with LIGO/Virgo~\cite{PhysRevLettLIGO}.  It is likely that a future gravitational wave event will be detected from a neutron star binary system.   In this case, it is possible to consider a flux of axions from the neutron star, which might create a gamma-ray signature detectable by Fermi-LAT.  In Ref.~\cite{BerenjiAAS2017}, we derive the spectral energy distribution from an extended source with a temperature $T=20$ MeV.  The phase diagram for QCD matter~\cite{ruster2005phase}, could apply for neutron stars, allowing for temperatures of up to $T=90 $MeV.  We consider the direct coupling of up quarks ($u$) and down quarks ($d$) to axions through a two-loop QCD Feynman diagram.

Our previous search for axions from neutron stars depended on the axion-coupling to quarks via $NN$-bremsstrahlung, where according to the Lagrangian $\mathcal{L}$, the derivative couples to the axion field as 
\begin{equation} \mathcal{L} \subset \frac{1}{f_a} g_{aNN}\left(\partial_\mu a\right) \bar{N} \gamma^\mu \gamma_5 N, \end{equation}
c
	where $g_{aNN}$ is the axion-nucleon-nucleon coupling, $a$ is the axion field, $N$ is the nucleon field, $\gamma_\mu$ represents the gamma matrices $\gamma^1,\gamma^2,\gamma^3$, and $\gamma_5=i\gamma^0\gamma^1\gamma^2\gamma^3$.
	However, in the model of QCD phase of matter for neutron stars, it may be more appropriate to study the axion coupling to quark via the triangle diagram and the chiral anomaly mediated by gluons.  The axion to gluon coupling may be characterized as                                                                                                                                                                                                                             \begin{equation} \mathcal{L} \subset \frac{a}{f_a} G_{\mu\nu} \tilde{G}^{\mu\nu} . \end{equation}

	Using the axial vector current, we obtain the quark couplings to axions $g_{au\bar{u}}$ and $g_{ad\bar{d}}$~\cite{graham2013}.  It will be shown that the axion coupling to quarks can prove to be more sensitive to lower masses than the nucleon coupling to axions, and to satisfy the criteria of ultralight axions (ULA).  ULA are postulated to be a relevant form of  dark matter~\cite{ULA}. 

	The paper is organized into sections as follows.  In Section~\ref{sec:AxionCurrent}, we discuss the theoretical implications of the axion current using triangle diagrams with coupling to quarks.  In Section~\ref{sec:APmodel}, we describe a theoretical model for calculating the emissivities.  In Section~\ref{sec:pred}, we discuss predictions for emissivities and spectral energy distributions for various $q\bar{q}$ channels.  In Section~\ref{sec:disc}, we present the significance of this work, the implications of the theoretical sensitivities to axions, and potential for gamma-ray experimental constraints based on the models described herein. Throughout this work, comparisons are also made between gamma-ray and gravitational-wave sensitivities.

\section{Axion Current}\label{sec:AxionCurrent}

%\documentclass[12pt]{article}
%\usepackage{physics}
%\usepackage{feynmp-auto}

%\newcommand{\ra}{\frac{q^2}{m^2}}
%\begin{document}
%\title{Axion coupling to quarks}

\begin{figure}
	\begin{centering}
		\includegraphics[width=12cm,clip,trim=0 1.5cm 0 0.2cm ]{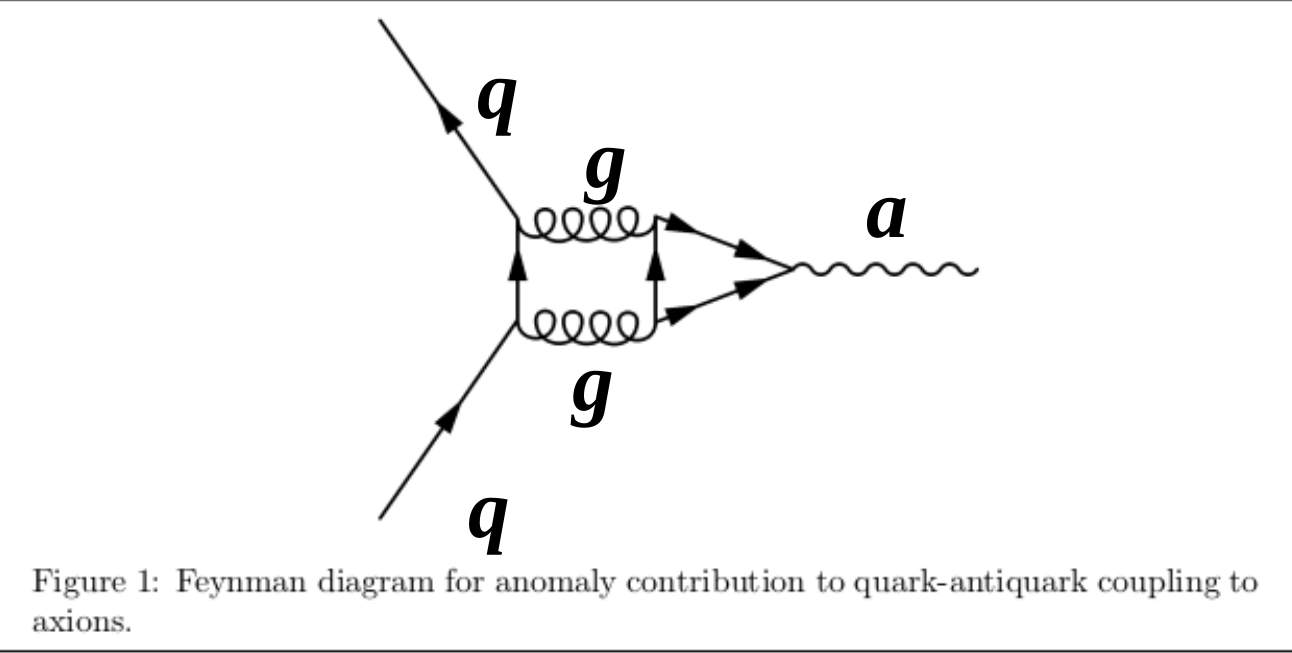}
		\caption{Anomaly diagram for anomaly contribution to quark-antiquark coupling to axions.}
		~\label{fig:feyn}
	\end{centering}
\end{figure}

\begin{comment}
\begin{figure}[h][t]
	\begin{mdframed}
	\begin{centering}
\begin{fmffile}{b}
\begin{fmfgraph}(200,150)
\fmfleft{i1,i2}
\fmfright{o}
\fmf{fermion,label=$Q$}{i1,v1}
%\fmf{phantom}{v1,v3} % Invisible rubber band
\fmf{fermion,label=$\bar{Q}$}{v2,i2}
%\fmf{phantom}{v2,v4} % also invisible rubber bando
%\fmf{fermion}{v1,v4}
\fmf{gluon,label=$g$}{v1,v2}
%\fmf{gluon,label=$g$}{v2,v4}
%\fmf{fermion}{v1,v2}
%\fmf{fermion}{v3,v4}
%\fmf{fermion}{v3,v5}
%\fmf{fermion}{v4,v6}
%\fmf{fermion}{v5,o1}
%\fmf{fermion}{v6,o2}
%\fmf{phantom}{v1,v2}
%\fmf{phantom}{v3,v4}
\fmf{fermion}{v1,v7}
\fmf{fermion}{v2,v7}
\fmf{boson,label=$a$}{v7,o}
%\fmf{phantom}{i2,v4,o}
%\fmf{phantom}{i1,v3,o}
%\fmf{phantom}{v5,v3}
%\fmf{phantom}{v6,v4}
%\fmf{phantom}{v8,o}
%\fmf{phantom}{v8,o}

	%\begin{fmfgraph}(40,25)
	%	\fmfleft{i1,i2}
	%	\fmfright{o1,o2}
%\fmf{fermion}{i1,v1,o1}
%/\fmf{fermion}{i2.v2,o2}
%\fmf{photon}{v1,v2}
%\fmf{gluon}{v1,v3}
%\fmf{gluon}{v2,v4}
%\fmf{fermion}{v3,v4}
\end{fmfgraph}
\end{fmffile}
\caption{Feynman diagram for one-loop quark coupling to axions.}\label{fig:feyn1}

\end{centering}
\end{mdframed}
%\end{mdframed}
\end{figure}

\end{comment}

The $q\bar{q}$ couples to axions via a chiral or Adler-Bell-Jackiw (ABJ) anomaly diagram with two loops, as shown in Figure~\ref{fig:feyn}.  Elementary discussions have been given in Ref.~\cite{chengLi}, the latter of which discusses at sufficient detail, the heavy quark effective theory (HQET).  %It is also possible to consider the one-loop coupling in Figure~\ref{fig:feyn1} 
We consider the two-loop coupling due to the axial current and the chiral anomaly, in order to improve the accuracy at gamma-ray energies.  
The current describing the quark to axion coupling has been described as~\cite{Srednicki1985}
\begin{equation} J_\mu = f_a \partial_\mu a + \frac{1}{2}u\gamma_\mu\bar{u} - \frac{1}{2}d\gamma_\mu\bar{d} + J^5_\mu \end{equation}

where the axial vector current may be written as~\cite{Srednicki1985}

\begin{equation} J^5_\mu = \frac{1}{2}\bar{u}\gamma_\mu\gamma_5 u + \frac{1}{2}\bar{d}\gamma_\mu\gamma_5 d \end{equation}\label{eq:axialVectorCurrent}
	under assumption of pseudovector coupling.  
		\subsection{Chiral Anomaly}
		The chiral anomaly arises from the Adler-Bell-Jackiw (ABJ) equation when applied to QCD.   The ABJ equation was originally written as a description of anomalies in electromagnetism, involving the electromagnetic tensor $F_{\mu\nu}$.  
		The chiral anomaly is invoked to explain the non-conservation of the axial vector current~\cite{PeskinQFT}.  In what follows, $\alpha_s$ is the strong force coupling.  The non-vanishing contravariant derivative of the current in Eq.~\ref{eq:axialVectorCurrent} is given by	
		\begin{equation} \partial^\mu J_\mu^5 = 2im_Q\bar{Q}\gamma_5Q + \frac{\alpha_s}{4\pi}T_R G\tilde{G} \end{equation}\label{eq:ABJ}
			where $Q$ represents the quark, $T_R=\frac{1}{2}$, and $G$ is the gluon field-strength tensor.  There are two contributions to Eq.~\eqref{eq:ABJ}: the first term is due to the non-zero quark mass, which arises from the spontaneously broken symmetry; the second term is the ABJ equation for QCD~\cite{PeskinQFT}.  We may assume the form for $J^5_\mu$ in Eq.~\ref{eq:axialVectorCurrent}.  The derivation of the form factor in the following subsection, in the model of axions from quark matter that we present in this paper, arises from taking the matrix elements. In the context of QCD, as we discuss here, ABJ is a QCD perturbation in powers of $\alpha_s$.The chiral anomaly of QCD and the spontaneously broken symmetry are invoked, by the ABJ equation, to explain the non-conservation of the axion current. 
		\subsection{Form Factor}
		In order to obtain the theoretical emissivities, luminosities, and energy fluxes for axions from neutron stars, we need to obtain the form factor, thereby computing the axion-quark coupling.  In our calculations, we may consider the customary momentum transfer
\begin{equation} \vec{q} = \vec{p}_1 - \vec{p}_2,  \end{equation}
	and the customary center-of-mass (CM) energy
	\begin{equation} s = \left(\vec{p}_1 + \vec{p}_2\right)^2 \end{equation}

		The vertex function, $\Lambda_{Q,\mu}$ may be expressed in terms of the form factors~\cite{Bernreuther}, which are functions of the momentum transfer $q^2$, as shown below  
		\begin{equation} \Lambda_{Q,\mu} = \gamma_\mu \gamma_5 F_1^5(q^2) + \frac{1}{2m_Q} p_\mu \gamma_5 F_3^5(q^2). \end{equation}
		The form factor $F_3^5$ can be shown to be complex in general.  We may consider the case $q^2\gg 4m^2$.  In Ref.~\cite{Bernreuther}, the renormalization has been taken into account in the derivation of the form factors. 
		The real part of the form factor $F_3^5$~\cite{Bernreuther} is
%\begin{alin}
\begin{equation} \Re F_3^5 (q^2) = \frac{\alpha_s^2}{32\pi^2}\left(\frac{2}{\left(q^4\right)}\right)\left[\log^2\left(\ra\right)-6\log\left(\ra\right) -2\zeta(2)\right)
		 + \frac{2}{\left(\ra\right)^2}\left[-32\log(\ra)+8\zeta(2)+12\log^2\left(\ra\right)\right]. \end{equation}
%	\end{align} 
		The imaginary part of the form factor may be written as~\cite{Bernreuther}: 
		\begin{equation} \Im F_3^5 (q^2) = \frac{4/\pi}{q^4}\left[3-\log\left(\ra\right)\right] \\  +\frac{1/\pi}{\left(q^4\right)}\left[32-24\log\left(\ra\right)\right]\end{equation} 
	We may in general consider the amplitude $|F_3^5(q^2)|$.  The axion-quark coupling can be written, in a form which uses the form factors, from invoking the Goldberger-Treiman relation~\cite{PeskinQFT}:

		\begin{equation} g_{aQ\bar{Q}}(q^2) = \frac{1}{2f_a}q^2\left|F^5_3(q^2)\right|.  \end{equation}
			
				The derivation is completely analogous to the pion-nucleon coupling via a triangle diagram.  
		
The tree-level s-channel (via gluon exchange) coupling can be safely neglected in considering the coupling of quarks to axions.  The form factor for the tree-level coupling can be written as
\begin{equation} \left| F_3^5\right| = \frac{1}{q^4} f_a^2 \alpha_s^4 \end{equation}
	We take $q=100$ MeV, $f_a\simeq 100$ GeV, $\alpha_s\simeq 10^{-2}$, $F_3^5\simeq 10^{-18}$ for tree-level coupling.  In this case, $F_3^5\left|_{\rm tree}|\right.\ll F_3^5\left|_{\rm two-loop}\right.$.  We do not consider the one-loop coupling, because it concerns the axial vector current.  
The tree-level coupling is proportional to $\alpha_s^{2}$, whereas the two-loop coupling is proportional $\alpha_s^{4}$.  The form-factor for the two-loop coupling may outweign the dependence on $\alpha_s^{4}$, such that the $\alpha_s^{2}$ term may be safely neglected. Note that $\alpha<1$.
	
	\subsection{Axion Emissivity of Quark Matter}\label{sec:emissivity}
	 The emissivity is a useful way to quantify the energy emitted by axions from neutron stars, as well as an intermediate calculation to calculate the flux, gamma-ray SED, and gamma-ray energy flux.  We can derive the axion emissivity proceeding first from the axion-quark coupling, which can be written by using the form factors and by invoking the Goldberger-Treiman relation~\cite{PeskinQFT}:

		\begin{equation} g_{aQ\bar{Q}}(q^2) = \frac{1}{2f_a}q^2\left|F^5_3(q^2)\right| \end{equation}
			
		Expressed in terms of the coupling to $g_{aq\bar{q}}$, we may write the volume emissivity $\epsilon_a$ of the quark matter to axions as a phase-space integral over the quark momenta:
	\begin{align} \epsilon_a &= \frac{1}{n}  \int_0^{p_F}  d\Omega_1 dp_1 \ p_1^2 \int_0^{p_F} d\Omega_2 \ dp_2 \ p_2^2 \\ &\times\int d\omega \omega^2 \ g^2_{aQ\bar{Q}}\left(q^2\right)^2 f_1\left(\omega;\mu,T\right) f_2\left(\omega;\mu,T\right) \delta(\omega - \sqrt{p_1^2 + m_Q^2} - \sqrt{p_2^2 + m_Q^2}). \end{align}  %\end{document}
		where $\omega$ is the quark energy, $p_{F,Q}$ is the quark Fermi momentum, and $f_i(E;\mu,T)$ is the Fermi-Dirac distribution for the quark chemical potential $\mu$ at neutron star temperature $T$.

		We may use the energy-time uncertainty relation to derive a timescale for the process of axion emission:
		\begin{equation} \Delta t \simeq \frac{1}{\Delta E} . \end{equation}

We may invoke this uncertainty relation, as the neutron star of quark matter is a Fermi-Dirac system of a quantum phase.  Further, the timescale for axion emission of 30 ms is of order of the timescale of the binary inspiral, and is thus relevant for the current study, whereas the gravitational wave signal lasted approximately 100 s.  
	Let us consider the defined quarks in the neutron star core. 	If we may na\"ively take the position uncertainty $\Delta x = 0.1$\ fm, then we have $\Delta p\simeq \times 10^{-13}$ MeV.  Assuming energy-time uncertainty relation, one may obtain a time uncertainty of 0.030 s for the processes in the medium.  Although axions would be produced quite quickly, the detection is not instantaneous.  The NS has to be close enough, and the detection efficiency is not high enough.  

		\begin{equation} f_i(E;\mu,T) = \frac{1}{1 + \exp\left(\frac{E-\mu}{T}\right)} \end{equation}
		We evaluate the width for the $q\bar{q}$ state, to a first approximation, as~\cite{PeskinQFT}:	
		\begin{equation} \Gamma(q\bar{q}\to a) = \frac{16\pi^2\alpha_s^2}{m_a^2} \left|\psi(0)\right|^2 \end{equation}	
		where $\psi$ represents the axion wavefunction.  
		%\begin{comment}
			\begin{figure}
			\begin{centering}
				\includegraphics[width=10cm]{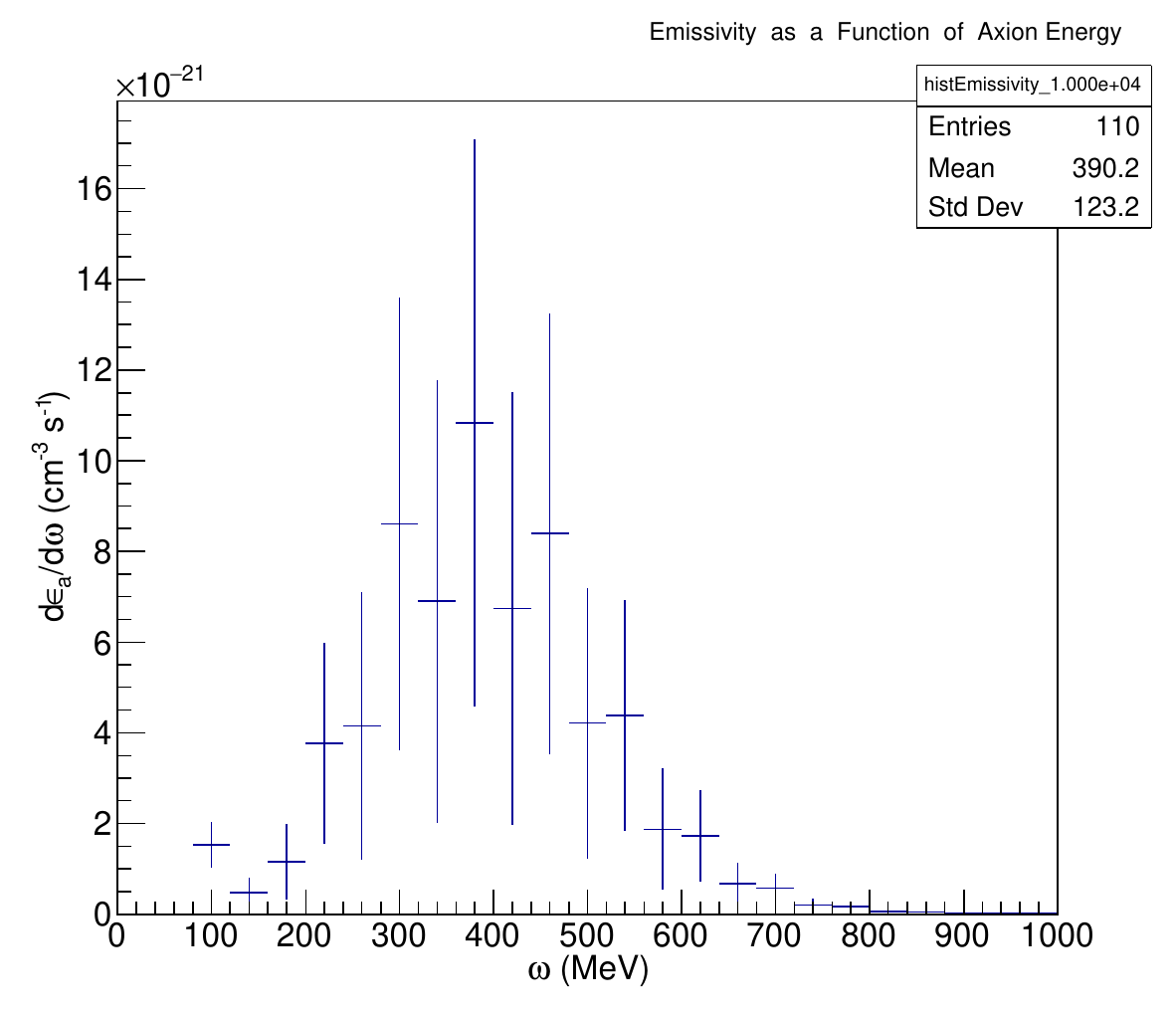}%emissivity_T100MeV.pdf}
				\caption{Emissivity as a function of axion energy for $T=100$ MeV}
			\end{centering}
		\end{figure}
	%	\end{comment}

		The axion-quark-antiquark coupling can be written in terms of the Goldberger-Treiman relation:
		\begin{equation} g_{aq\bar{q}} = \frac{q^2}{2f_a}\left|F_3^5(q^2)\right| \end{equation}
		The cross-section amplitude may be written in the form:
\begin{equation}
\sigma(q\bar{q}\to a) = \frac{12\pi^2}{M}\Gamma(a\to q\bar{q})\delta(M^2-s). \end{equation}

	%\begin{figue}[h][t]
	%		\begin{centering}
	%			\includegraphics{fig2a}%emissivity_T100MeV.pdf}
	%			\caption{Emissivity as a function of axion energy for $T=100$ MeV}
	%		\end{centering}
	%	\end{figure}

\begin{comment}
		The spectral energy distribution, assuming a point source model, can be related to the emissivity as follows~\cite{berenjiAxions}:
		\begin{equation} E\frac{d\Phi}{dE} = \frac{d\varepsilon_a}{d\omega} V_{NS} \Delta t \frac{\Gamma_{a\gamma\gamma}}{4\pi d^2}, \end{equation}
			where $V_{NS}$ is the neutron star volume, $d$ is the distance to the source, and $\Gamma_{a\gamma\gamma}$ is the partial width of axions to two photons.
\end{comment}

	\section{Astrophysical Model}\label{sec:APmodel}

	\subsection{Quark Matter Neutron Star}
In the model that we consider, the neutron star is may be considered to be in a QCD phase of quark matter, or a mixed phase of hadronic and quark matter~\cite{Weber}.  This phase of quark matter is considered to be color superconducting.  In addition, it is possible to consider the ``CFL'' phase where strange quarks, in addition to up and down quarks, participate in Cooper pairing mechanism~\cite{Alford2009}.  This phase is superfluid and exhibits broken chiral symmetry.  The neutral quark (NQ), the gapless CFL, and the gapless 2SC phases are candidates for stellar matter~\cite{Ruster2005neutrino}.  

We may assume a number density of $n=0.15 $fm$^{-3}$ within the neutron star/quark star and a quark chemical potential of $\mu=400$ MeV in Ref. ~\cite{ruster2005phase}.  We also cite the more recent reference of 2014 ~\cite{Hassaneen}, which yields a similar density of $\rho\simeq$0.15fm$^{-3}$. We assume a quark matter or a mixed quark-hadron phase for the neutron star.   We assume a Fermi momentum $p_F$ which is given by:

\begin{equation} p_F =  \left(3\pi^2n\right)^{1/3} \end{equation}

which is calculated to be 323.81 MeV from the density that we cite.  

The temperature of the QCD phase of quark matter depends on the exact description of the neutron star in the QCD phase diagram.  In deriving the prediction for gamma-ray signal below, we assume $T=90$ MeV.

\subsection{Calculation of the Photon Flux}\label{sec:photonflux}

The photon flux can be derived in a stepwise procedure in a Monte Carlo simulation, largely relying on the results of Section \ref{sec:emissivity}.
First, we integrate over momentum space.
\begin{equation} \mathcal{D}v = d^3p_1 d^3p_2 \frac{\omega^2}{[1+e^{(E-\mu)/T}]^2}g_{aqq}^2 \end{equation}
Next, we define the neutron star volume, assuming a radius of 10 km.
\begin{equation} V_{NS} = \frac{4\pi}{3}\left(10^6\right)^3 \ \mbox{cm}^3 \end{equation}
We provide a delta function, with respect to the kinetic energy $\omega$, invoking the $u$ and $d$ momenta and masses, $m_u$ and $m_d$.
\begin{equation} \delta\left(\omega - \sqrt{p_1^2+m_u^2}-\sqrt{p_2^2+m_d^2}\right) \end{equation}
The axion-quark-quark coupling is derived in terms of the real and imaginary components of the form factor $F_3^5$,
\begin{equation} g_{aqq} = \frac{1}{2}\sqrt{\Re(F_3^5)^2 + \Im(F_3^5)^2}\frac{q^2}{f_a} \end{equation}

We write the axion wavefunction as follows.
\begin{equation} \psi_0 = 10^{-10} {\mbox cm}^{-3/2} \end{equation}
We derive the following variable $\lambda$ as such:
\begin{equation} \lambda = \frac{16\pi}{3}\alpha_s^2 \left(\psi_0^2\right)^{-1}\end{equation}
The cross section is written as shown below.
\begin{equation} \sigma = 12\pi^2\frac{\lambda}{\sqrt{s}} \end{equation}
The derived derivative is defined in terms of $dv$ and the number density of quarks, $\rho$.
\begin{equation} \mathcal{D}\epsilon = \sigma\frac{\mathcal{D}v}{\rho^2} \end{equation}    
Finally, we obtain the differential photon flux:
\begin{equation} E\frac{d\Phi}{dE} = 2\frac{m_a^5}{4\pi d^2} \int_0^{p_F} \mathcal{D}\epsilon V_{NS} \lambda \Delta t \delta(E-\omega/2) \end{equation}

\section{Predictions from the Model}~\label{sec:pred}
\begin{figure}
	\begin{centering}
		\includegraphics[width=12cm]{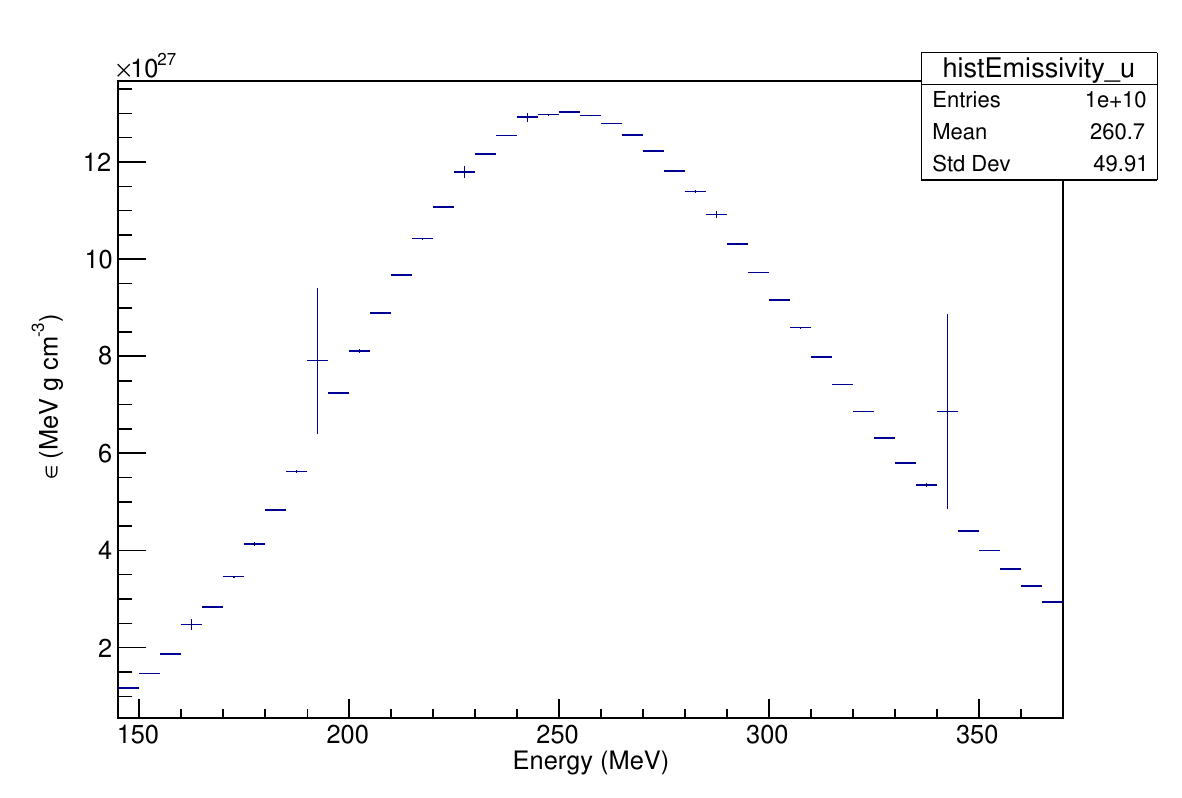}%emissivity_v14.pdf}
		\caption{Emissivity for axion energy loss from neutron star at $T=90$ MeV and assuming a mass of $m_a=1$meV. }
		~\label{fig:emiss}
	\end{centering}
\end{figure}

The calculated emissivity is within several orders of magnitude of the SN energy loss rate into neutrinos, which indicates some similarities which would be expected in the energy loss rate.  The emissivity as a function of axion energy is shown in Figure ~\ref{fig:emiss}.  The emissivity according to our model is quite large, even with a small axion mass of order $\mathcal{O}$(1 meV)

According to this model, we may demonstrate a peak in the spectral energy distribution (SED), with a sharp cutoff before the Fermi momentum, as shown in Figure~\ref{fig:SED1} for the $u\bar{u}$ axion coupling for $T=90$ MeV.   This is very near the Fermi-LAT point source sensitivity of 5$\times 10^{-9}$ cm$^{-2}$ s$^{-1}$.  

		\begin{figure}
			\begin{centering}
	\includegraphics[width=12cm]{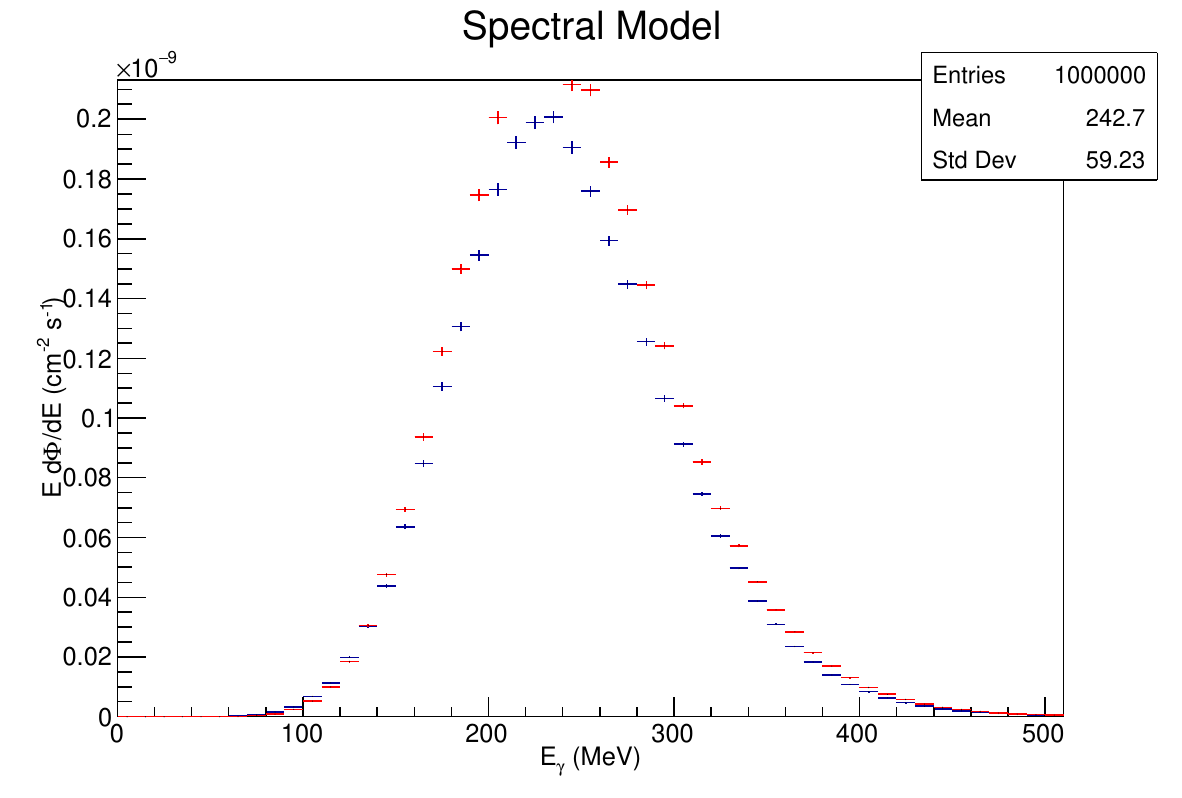}%Eflux_u_d_T90.pdf}
				\caption{Spectral Model for gamma rays produced from a neutron star at a distance of 100 pc and assuming an axion mass $m_a=1$ meV, for $u\bar{u}$ and $d\bar{d}$ channels. The expected axion emissivity from the neutron star model is plotted versus gamma ray energy.  }~\label{fig:SED1}
			\end{centering}
		\end{figure}

		\begin{figure}
			\begin{centering}
				\includegraphics[width=12cm]{fig5.pdf}%plot_Emissivity_new18_1e-11.png}
				\caption{Spectral model for axion emission, including both $u\bar{u}$ and $d\bar{d}$ channels.}~\label{fig:sed_ud}

			\end{centering}
		\end{figure}
	
		\begin{figure}
			\begin{centering}
				\includegraphics{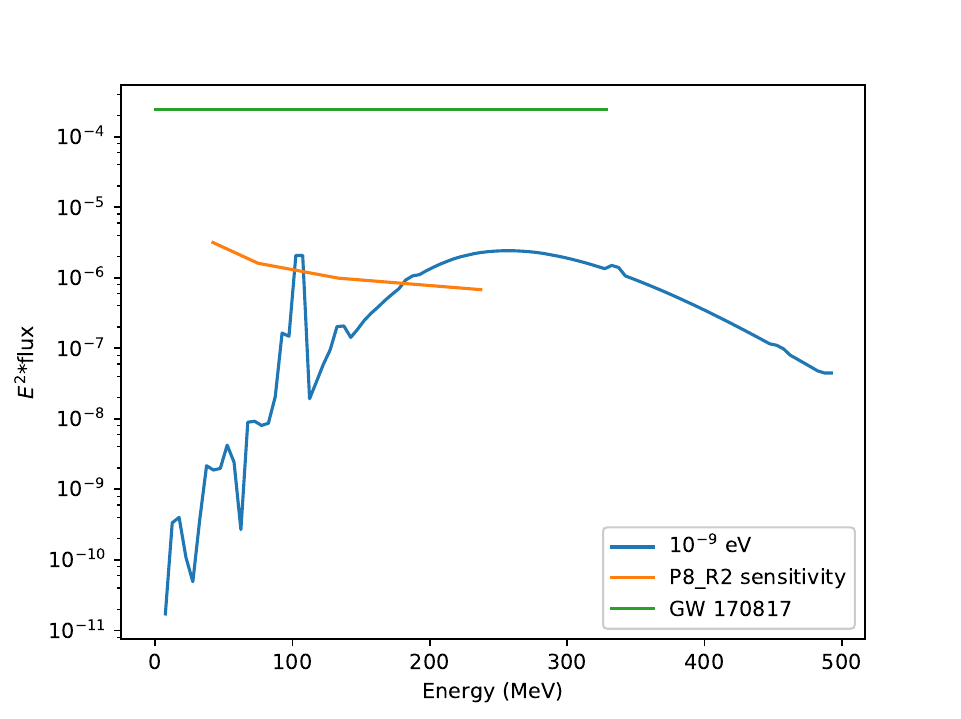}%histE2flux_ma_new11_1e-9.pdf}
				\caption{Estimated sensitivity of detecting axion signal with Fermi-LAT, for various values of $m_a$, for a binary neutron star at a distance of 1 Mpc.}~\label{fig:sens}
			\end{centering}
	\end{figure}

		\begin{figure}
			\begin{centering}
				\includegraphics{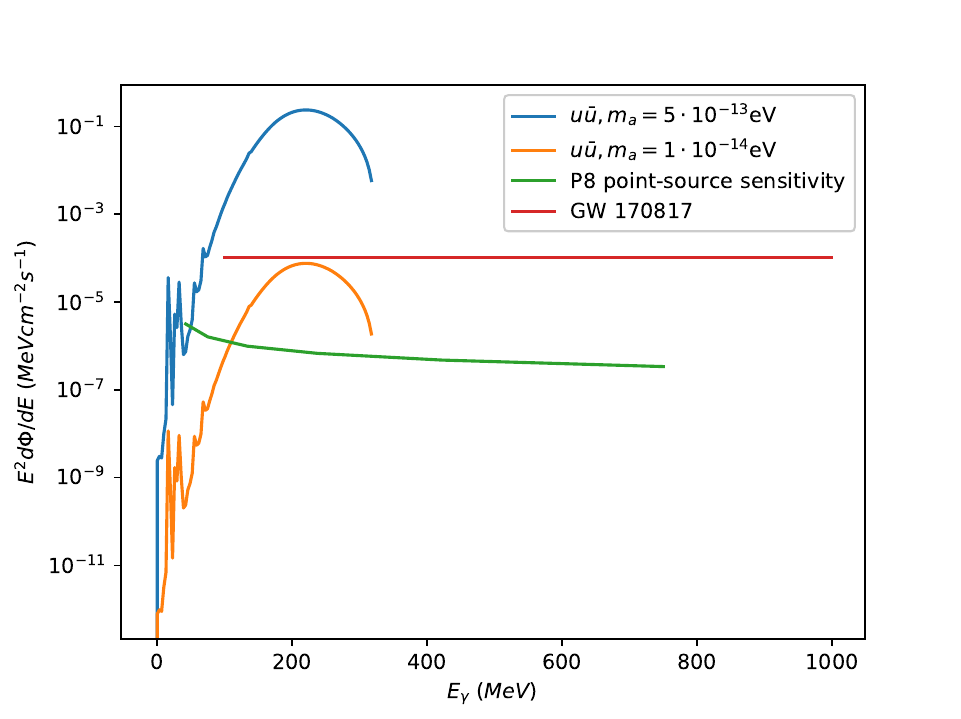}%sens_plot_new6.pdf}
				\caption{Estimated sensitivity of detecting axion signal with Fermi-LAT, for various situations at a distance of 100 kpc.}~\label{fig:sens_plot}
			\end{centering}
		\end{figure}

		The spectral energy distribution for combined $u\bar{u}$ and $d\bar{d}$, as shown in Figure~\ref{fig:sed_ud} is within the range of experimentally measurable flux with Fermi~\cite{instrumentPaper}.  
		It may be possible to set constraints on the axion mass by setting upper limits on the signal, as shown in Figure ~\ref{fig:sens}, where the spectral model is compared to the 10-year point source sensitivity of the Fermi LAT. At a distance of 100 kpc, the Fermi LAT could detect the signal arising from the NS-NS merger, from a simple inverse-square law calculation.    

%\listoffigures

\begin{table}
\begin{tabular}{|l|l|}
\hline
variable/symbol & meaning \\
\hline
$a$ & axion particle \\
$\alpha_s$ & strong force coupling (dimensionless)\\
$\epsilon$ & emissivity \\
$F_3^5 $ & form factor \\
$f_a$  & axion decay constant \\
$\Gamma$ & decay width \\
$m_a$ & axion mass \\
$q^2$ & momentum transferred squared \\
$q\bar{q}$ & quark/anti-quark state \\
\hline
\end{tabular}
\caption{Table of variables and usage in this article.}
~\label{tab:variables}
\end{table}

\section{Discussion} ~\label{sec:disc}

%\section{Predictions from the Model}
\begin{figure}
	\begin{centering}
		\includegraphics[width=12cm]{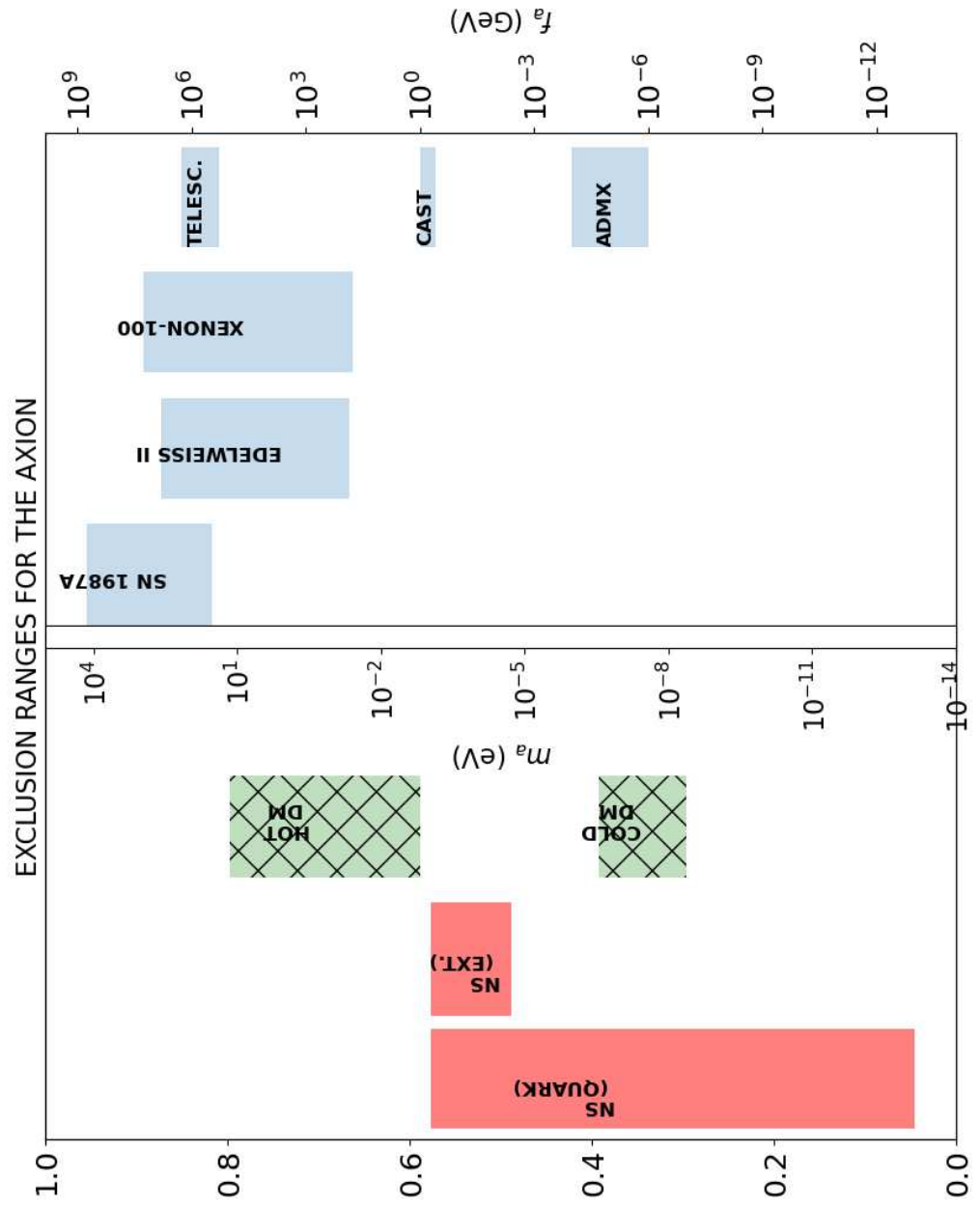}%emissivity_v14.pdf}
		\caption{Exclusion plots for various experiments and axion decay channels.  Solid rectangles refer to an excluded range; hash-filled rectangles refer to an included region.  Upper and lower limits are the basis for this diagram.  Limits in shades of red represent those derived from axion decay modes. See Refs.~\cite{art50,art51,art53,art54,art55}}.
		~\label{fig:fig8}
	\end{centering}
\end{figure}

The Fermi LAT was designed to measure gamma-rays of energy of 30 MeV, but in reality, this was not possible.  Further, a multi-messenger study involving gamma-rays and X-rays would improve upon a Fermi gamma-ray study alone, especially for lower temperatures of the neutron stars.  Fermi-LAT pointed observations of a NS-NS merger, triggered by LIGO, would improve the recent limits from LIGO GWS 170817. 
In Figure 2, we plot the emissivity of axions, under the scenario discussed in this article, and find that the mean energy is 390 MeV.  In the context of Figure 3, this makes sense, because we find for figure 3 that the median energy is $~$240 MeV.  We plot meV and sub-neV for the emissivities and fluxes for NS under merger conditions.  We determine that sub-neV for $m_a$ is favored, because it is the lowest mass consistent with the physical conditions.  In Figure 6, the axion is only marginally detectable by Fermi-LAT and LIGO/Virgo. In Figure 7, however, the axion is detectable (within the margin of error) by both experimental teams.

If an improved gamma-ray telescope were developed, what would be desired would be detection of gamma-rays down to 5 MeV.  Further, if a multi-messenger campaign were commissioned, an X-ray telescope would be complementary, measuring X-ray photons down to lower energies.
The model of neutron star quark matter that we rely on here, Ref.~\cite{Ruster2005neutrino}, is not the only one, and a study was done in Ref.~\cite{Hujeirat}.  However, Ref.~\cite{Ruster2005neutrino} seems to be more thorough. 

Figure~\ref{fig:fig8} highlights the results of the current work in a spectacular fashion.  There is no other limit which pushes down into the mass parameter space as much as this one.  There is no other limit which excludes the dark matter parameter space as much as this one, except for ADMX perhaps.  From the limit presented here, we exclude axions as Hot Dark Matter and allow it to be Cold Dark Matter.  

\section{Acknowledgments}

Many thanks to the wonderful people of SLAC National Accelerator Laboratory, namely Prof. Michael Peskin of the Theory Group, who gave me some valuable comments on this article; and who taught me quantum field theory in the first place when I was a graduate student at Stanford University.  I wish to acknowledge the Fermi Large Area Telescope collaboration classification of this paper as a Cat-III paper.  Thanks for some funding from California State University, Los Angeles, for which credit may acknowledged to  to the former Chair of the Physics Department, Prof. Radi Jishi.  Thanks to Dr. Riccardo DeSalvo of the LIGO Scientific Collaboration for the useful discussions on multimessenger astrophysics.

%\input{JMP_article_berenji_NSmerger_v4.bbl}%JMP_article_berenji_NSmerger_v4.bbl

%\begin{comment}
%\begin{thebibliography}
%\bibliography{refs4.bib}

%\end{thebibliography}
%\end{comment}
%\input{refs3old.bbl}
%\end{document}

%\input{physRevC_berenji_3.bbl}
	%	\bibliographystyle{plain}
	%	\bibliography{refs3old}
\end{document}